\documentclass[twocolumn,prl,aps,preprintnumbers,amsmath,amssymb
,floatfix]{revtex4-1}   

\usepackage{amssymb}
\usepackage{amsmath}
\usepackage{graphicx}
\usepackage{textcomp}
\usepackage{color}
\usepackage{amsfonts}
\usepackage{epsfig}
\usepackage{hyperref}

\newcommand{\arctanh}[1]{\operatorname{arctan}}

\begin{document}

\title{Disorder and dephasing effect on electron transport through conjugated molecular wires 
in molecular junctions} 
\date{\today} 

\author{Daijiro Nozaki$^{1}$}
\author{Claudia Gomes da Rocha$^{1}$}
\email[Corresponding Author, Electronic address:]{cgomes@nano.tu-dresden.de}
\author{Horacio M. Pastawski$^{2}$}
\author{Gianaurelio Cuniberti$^{1,3}$}
\affiliation{$^{1}$Institute for Materials Science and Max Bergmann Center of Biomaterials, 
TU Dresden, 01062 Dresden, Germany} 
\affiliation{$^{2}$Instituto de Fis\'{i}ca Enrique Gaviola (CONICET) and FaMAF, Universidad Nacional de C\'{o}rdoba Ciudad Universitaria, 5000  C\'{o}rdoba, Argentina}
\affiliation{$^{3}$Division of IT Convergence Engineering and National Center for
Nanomaterials Technology, POSTECH, Pohang 790-784, Republic of Korea}

\begin{abstract}
\noindent
Understanding electron transport processes in molecular wires connected between contacts is a central focus in the field of molecular electronics. Especially, the dephasing effect causing tunneling-to-hopping transition 
 has great importance from both applicational and fundamental points of view. 
   We analyzed coherent and incoherent electron transmission through conjugated molecular wires by means of density-functional tight-binding theory within the D'Amato-Pastawski model.  Our approach can study explicitly the structure/transport relationship in molecular junctions in a dephasing environmental condition using only single dephasing parameter. We investigated the length dependence and the influence of thermal fluctuations on transport  and reproduced the well-known tunneling-to-hopping transition. 
   This approach will be a powerful tool for the interpretation of recent conductance measurements of molecular wires.
\end{abstract}


\maketitle


\section{Introduction} 
The study of electron transport through nanostructures has been a primary interest in the field of molecular electronics in the last few decades.
Recently, increasing attention has been paid to the main role played by phase-breaking processes in molecular junctions. \cite{Maassen2009,Zilly,Cattena,Mdey,Nozaki} The transition from coherent (tunneling) to incoherent (hopping) regimes is one example of such a process which provides helpful information for the fundamental understanding of transport mechanisms and for applications such as organic semiconductors. 
Such transitions have commonly been reported in many studies of electron transfer in molecular systems and  electron transport in molecular junctions, for instance,  self-assembled monolayers, \cite{1482,Frisbie11} polymers, \cite{Frisbie11,1482,WBD,Hines,Tada,Selzer} and macromolecules such as proteins, \cite{refpro} DNAs, \cite{refDNA} and organic semiconductors. \cite{WBD}

For further development of   molecular devices, 
the intrinsic mechanisms of charge transport including the dephasing effects and influential factors such as thermal fluctuations need to be understood in detail. 
In order to evaluate the electronic conductance of nanostructures, the Landauer formula is commonly used. 
However, this method is restricted to address systems within the coherent regime. 
In principle, this method must be extended before it can be applied to organic semiconductors or disordered polymers where the dephasing effects play an essential role in electron transport. 
%
Normally, many molecular aggregates widely used in organic semiconductor devices exhibit high
disorder features. A proper theoretical description of the transport properties of such defective structures 
has to take into account not only band-like coherence but also dephasing mechanisms via a hopping representation.  \cite{Troisi}

One of the main sources of dephasing processes is the interaction between electronic states and molecular vibrations,  referred as electron-phonon (\textit{e-ph}) coupling. 
The \textit{e-ph} interactions can be experimentally inferred from inelastic tunneling spectroscopy 
or from thermal resonance broadening.
 The challenge, then, concentrates in incorporating \textit{e-ph} interactions into the 
Landauer approach. 
The current state of the art as followed by several theoretical works involves the use of density functional theory (DFT) within nonequilibrium Green's function formalism where \textit{e-ph} interaction is modeled by self-energy terms
that can be solved self-consistently. \cite{refIETSDFT} However, such a strategy can be computationally demanding. 
Such limitations can compromise the understanding of the main role played by
dephasing effects on the charge transport properties of molecular devices, leading to the search for less expensive models that attend to more qualitative interpretations.
%
%
%
%
One example of such a model is  B\"{u}ttiker's scattering approach, \cite{refButtiker,Maassen2009} in which \textit{e-ph} coupling is replaced by phenomenological voltage probes capable of inducing phase-breaking process into the molecular conductor. 
%

Other approaches using reduced density matrix elements where their time evolution can be
described by Redfield equation have also been adopted to describe the physics of electron  transfer/transport in molecular
systems and their crossover trends between coherent and incoherent regimes. \cite{Felts1995,Segal2000,Segal20002}
%
%
%
%
Nevertheless, it is highly desired to investigate such a prominent physical phenomenon using a 
parameter-free (or using as few parameters as possible) first principles methodology since a more detailed quantitative study can be rendered using such techniques. 
%
Such tecniques would enable us to investigate the relationship between  the conformational change of a realistic molecule  and the change of electron transport.

In this letter, we analyzed the main role played by dephasing effects in the transport behaviour of organic polymer systems 
using the Hamiltonian version of B\"{u}ttiker's scattering approach which is the D'Amato-Pastawski model.  \cite{refDAmato}  We show that the conductance of the molecules can be dramatically  modified due to the dephasing effect depending on the size of the molecules. 
Important fundamental questions such as  the tunneling-to-hopping transition and influence of thermal fluctuation on transport  are also addressed in this work. 
We demonstrate that dephasing has a significant impact in driving the transport regime of molecular systems from a quantum mechanical tunneling to Ohmic behaviour  for longer molecular wires. 
%


%
\begin{figure}[ht!]
\begin{center}
\includegraphics[width=8.5cm,clip=true]{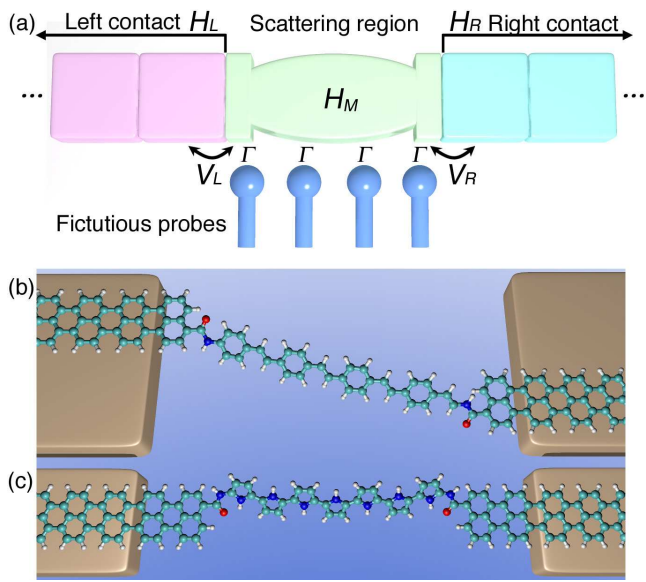}
\end{center}
\caption{\small{(a) A schematic model of a molecular junction connected between two electrodes.  B\"{u}ttiker probes are introduced to simulate dephasing events. (b) PPV- and (c) pyrrole-oligomer is coupled to electrodes consisted of graphene nanoribbons (GNRs) via peptide linkers.}} 
\label{fig1}
\end{figure}

\section{Theoretiacal framework} 
We calculated the conductance of molecular wires using the Landauer formula written in terms of Green's functions. \cite{gdatta} Figure \ref{fig1}(a) shows a schematic picture of the studied system which is composed of a two terminal conducting channel
connected to a central scattering region. 
A sequence of local dephasing probes 
 is incorporated to describe
incoherent events that affect the transport properties of the system. The electronic structure of the molecular wires is represented by a tight-binding Hamiltonian; 
   $H = H_\textrm{L}+V_{\textrm{L}}+H_\textrm{M}+V_{\textrm{R}} + H_\textrm{R}$,
where $H_{\textrm{L/R}}$ and $H_\textrm{M}$ represent the left/right electrode and central contributions, respectively. $V_{\textrm{L/R}}$ defines 
the coupling between source/drain electrodes and the molecular wire. 
 The electronic propagator for the coupled system is represented by a retarded Green's function defined as $G^\textrm{R}(E)=[(E+i\eta)I-H_\textrm{M}-\Sigma_{\textrm{L}}-\Sigma_{\textrm{R}}]^{-1}$, where $\Sigma_{\textrm{L/R}}$ are the self-energy elements which include
the influence of the contacts. 
    The conductance of a 1D molecule at low bias  in the coherent regime is then obtained via Landauer's formula, 
$G=G_0\,\textrm{Tr}[ G^\textrm{R} \Gamma_\textrm{L}G^\textrm{A}\Gamma_\textrm{R}]$ where 
$G_0$ is the quantum conductance unit $G_0=2\frac{e^2}{h}$ and $\Gamma_{\textrm{L/R}}$ represents the broadening function given by $\Gamma_{\textrm{L/R}}(E) = i[ \Sigma_{\textrm{L/R}}(E) - \Sigma_{\textrm{L/R}}^\dag(E) ]$.

To include dephasing events in the model, the D'Amato-Pastawski formalism is adopted.
Within this picture, one includes additional self-energy terms that account for the dephasing processes. This function is defined as $ G^\textrm{R}_\textrm{eff}(E)= [(E+i\eta)I - H_{\textrm{M}} - \Sigma_{\textrm{L}}  - \Sigma_{\textrm{R}}  - \sum_{n=1}^N\Sigma_{\textrm{B}, n}    ]^{-1}$, where $N$ is the number of reservoirs and $\Sigma_{\textrm{B}, n}$ are the self-energies due to B\"{u}ttiker probes; $\Sigma_{\textrm{B}, n} = i\gamma_{\textit{e-ph}}/2$ that is written in term of a dephasing parameter, $\gamma_{\textit{e-ph}}$, associated to the $e$-$ph$ coupling.
The effective conductance is given by $T_{\textrm{eff}} = T_{\textrm{L,R}} + \sum_{m,n=1}^{N} T_{\textrm{L},m}W_{m,n}^{-1}T_{n,\textrm{R}}$,
  where $T_{\textrm{L,R}}$ describes the coherent contribution and the subsequent term takes into account the incoherent events. 
$W$ is a Markov matrix formed by   transmission $T_{m,n}$ between the $m$-th and $n$-th probes, and the reflection $R_{m}=1-\sum_{n \neq m}^N{T_{m,n}}$, defined as  $W_{m,n} = \delta_{m,n} (1-R_{m}) -(1-\delta_{m,n})T_{m,n}$. The broadening caused by dephasing processes is defined as $\Gamma_{m}\equiv i[\Sigma_{\textrm{B}, m}-\Sigma_{\textrm{B}, m}^\dag]=\gamma_{\textit{e-ph}}$ which is related to the dephasing rate, $k=\Gamma_{m}/\hbar$.  
Although the dephasing strength $\gamma_{e-ph}$ can be evaluated by the Fermi Golden rule quantitatively from first principles, which are comparable to $k_{\mathrm{B}}T$, \cite{Cattena,newRef} it is also possible to 
 examine the influence of the dephasing events on charge transport by changing the rate of dephasing events. 

\section{Result and discussion}

\begin{figure*}[ht!]
\begin{center}
\includegraphics[width=12cm,clip=true]{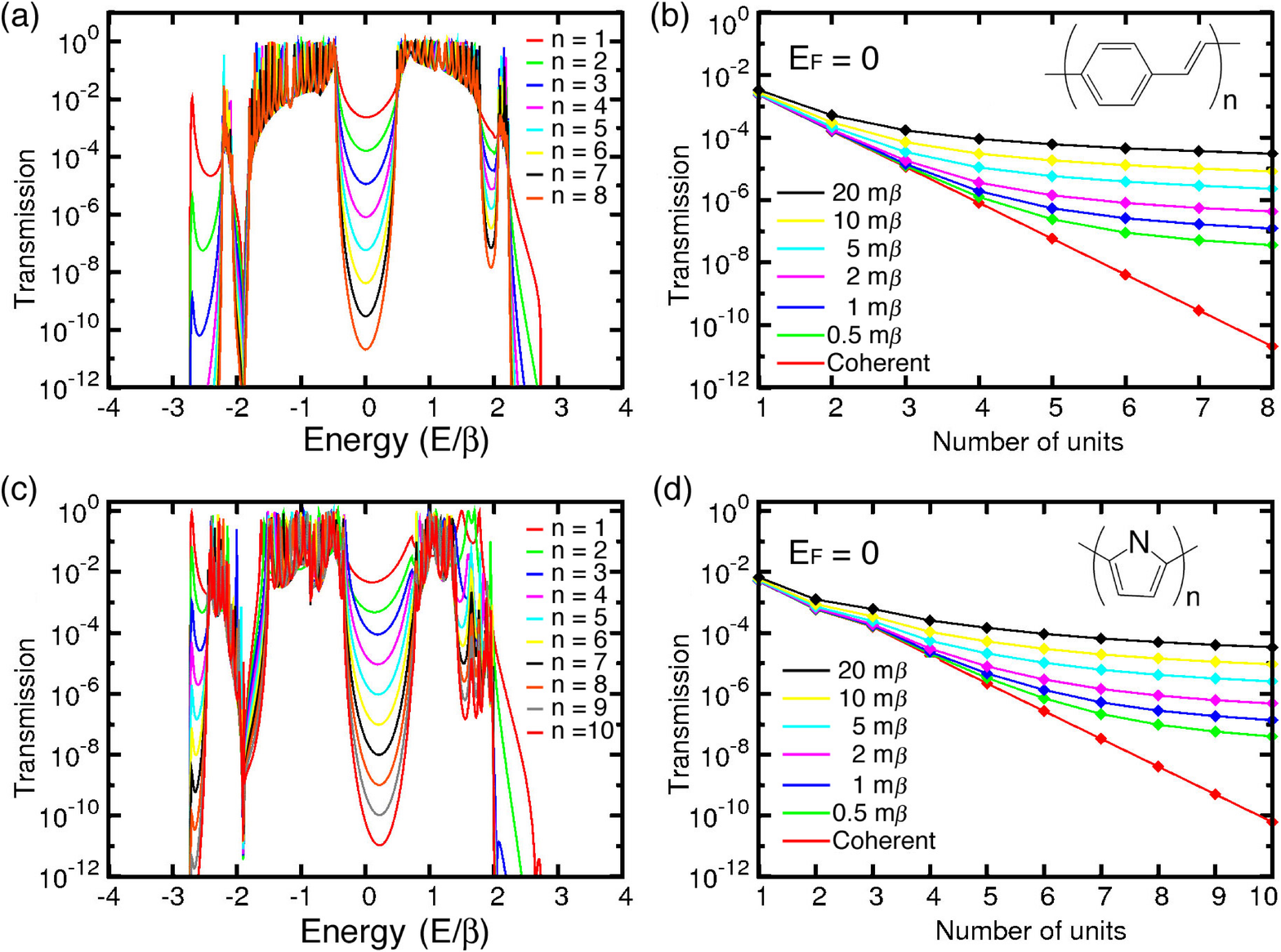}
\end{center}
\caption{\small{Electronic transmission calculated for PPV- (upper panels) and pyrrole- (lower panels) based molecular wires as a function of the molecular length within tight binding approximation. Results are displayed within coherent [(a) and (c)] and incoherent [(b) and (c)] regimes assuming also different dephasing parameters.}}
\label{fig2}
\end{figure*}

Firstly, it is important to obtain the electronic transmission within the coherent regime in order to compare it to the incoherent cases. We chose two sample systems for such analysis: PPV- [Fig.~\ref{fig1}(b)] and pyrrole-oligomer [Fig.~\ref{fig1}(c)] polymers. 
At first, we use a simple tigh-binding model, which only considers the nearest-neighboring $\pi$-orbitals in order to capture the trend of the transmission profiles. The coupling strength of C-C bonds on the GNR electrodes \cite{gnr1} are simply parameterized as $\beta=2.66$ eV. In the scattering region, the transfer integrals for double and single bonds are set as $\beta_{\textrm{d}}=1.2\beta$ and $\beta_{\textrm{s}}=0.8\beta$, respectively, and the on-site energies for each atom are $\epsilon_{2p}=$ 0.0$\beta$ (C), 0.902$\beta$ (N), and 1.902$\beta$ (O). \cite{Verkade} The Fermi energy of the systems are set to 0.0$\beta$ for simplicity. 
The coherent transmission results are displayed in Fig.~\ref{fig2} for (a) PPV- and (c) pyrrole-based molecular wires considering different lengths. The transport response of the systems is characterized by a transmission valley centered around the Fermi energy ($E_F=0.0\beta$). Such valleys become more pronounced as the molecular length increases and this results in the well-known exponential decay of the conductance at the Fermi level as shown in the logarithm plots of Fig.~\ref{fig2} (b) and (d) (red lines).

Dephasing effects are subsequently included in the model and we investigate how they affect the conductance behaviour as a function
of molecular length. Bearing in mind that the \textit{e-ph} coupling is much weaker than the transfer integrals, \cite{ep-coupling} we
finely changed the dephasing parameters between 0.5 and 20 m$\beta$ and studied their influence on the charge transport as shown in
Figs.~\ref{fig2} (b) and (d). One can see that the transmission is considerably enhanced in comparison to the coherent tunneling as the
length of the molecule increases. 
A clear deviation from exponential behaviour can be observed which becomes more evident with increasing
dephasing strength. 
\begin{figure*}[ht!]
\begin{center}
\includegraphics[width=12cm,clip=true]{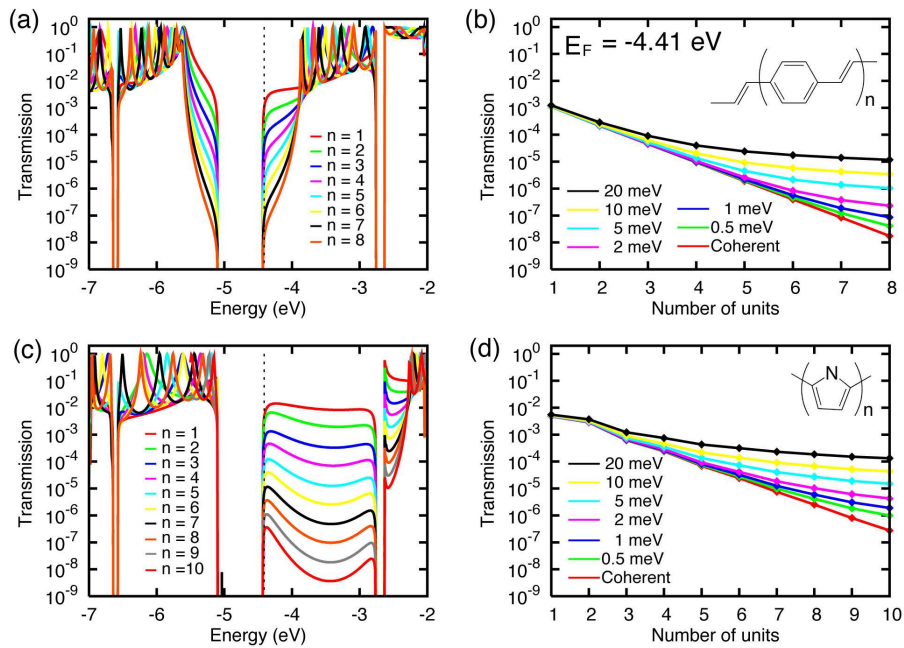}
\end{center}
\caption{\small{Electronic transmission calculated for PPV- (upper panels) and pyrrole- (lower panels) based molecular wires as a function of the molecular length within the DFTB approach. Results are displayed within coherent [(a) and (c)] and incoherent [(b) and (c)] regimes assuming also different dephasing parameters. }}
\label{fig3}
\end{figure*}

More sophisticated methods such as the density functional based tight-binding (DFTB) approach \cite{DFTB,DFTB-basis}
were also employed to verify the robustness of such crossover behavior. 
We relaxed the molecular junctions and calculated the transmission in the same way as ref \cite{switch}. At first, we relaxed
the isolated molecular wires and 1D GNRs individually. Then we coupled the relaxed
isolated molecular wires between the single unit cells of the relaxed 1D GNR
electrodes.  We relaxed them using periodic boundary conditions fixing the atoms at
the boundaries. After the relaxation, we placed the relaxed coupled systems between
the relaxed 1D semi-infinite GNR electrodes and calculated the transmission
probabilities using the same approach.

 The Fermi energy for the system was determined from the band
calculation of the 1D armchair GNRs which was determined to be -4.41 eV.  Because of
the steric repulsion between hydrogen atoms, the carbon-carbon distances connecting the unit cells in the
horizontal direction is longer than that of others. Therefore the armchair of GNRs
is slightly dimerized giving rise to the band gaps \cite{gnr2} around 5 eV, this can be seen as
abrupt drop in the transmission functions in Fig.~\ref{fig3}(a) and (c).
The transmission profiles and length dependence of conductance at the Fermi energy with/without dephasing effects calculated using DFTB method are presented in Fig.~\ref{fig3}. A similar
behavior as the one obtained via the simple tight binding approximation in Fig.~\ref{fig2} is observed. 
From the linear trend of coherent tunneling, we determined the damping
constants for PPV, 0.241\AA$^{-1}$, and pyrrole, 0.308\AA$^{-1}$. 
The exponential decay associated with coherent transport is quenched as the dephasing effects get stronger. 

\begin{figure}[ht!]
\begin{center}
\includegraphics[width=8.5cm,clip=true]{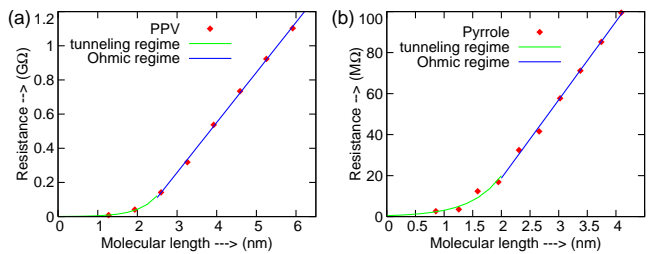}
\end{center}
\caption{\small{Resistance of (a) PPV- and (b) pyrrole-based oligomers as a function of molecular lengths. The fitting curves for coherent tunneling and hopping transport are shown in green and blue, respectively. Dephasing strength, $\gamma_{\textit{e-ph}}$, was set to 20 meV }}
\label{fig4}
\end{figure}

\begin{figure}[ht!]
\begin{center}
\includegraphics[width=8.5cm,clip=true]{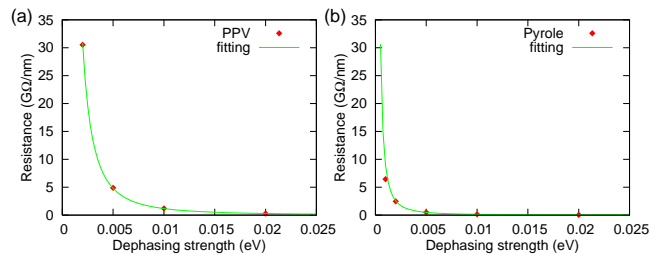}
\end{center}
\caption{\small{Resistance of (a) PPV- and (b) pyrrole-based oligomers per unit length in the Ohmic regime as a function of dephasing strength.  }}
\label{fig42}
\end{figure}

Once more, we verified that dephasing dictates the
conducting response for longer 1D systems. The change on the transport behavior from direct coherent tunneling to a hopping
mechanism can be evidently seen in Fig.~\ref{fig4} which shows the resistance as a function of length for both molecular structures calculated with the dephasing strength being 20 meV. The system
undergoes a crossover from the exponential to the Ohmic regime with the estimated resistance per unit length for PPV and pyrrole being
283.9M$\Omega$/nm and 39.3M$\Omega$/nm, respectively. 
Figure ~\ref{fig42} shows the resistance per unit length in the Ohmic regime as a function of dephasing strength for both molecular wires. We can see that the resistance is decreased because of the assistance from the incoherent transport.

Since we want to assess the 
influence of conformational fluctuations on 
transport in realistic molecular wires under dephasing conditions, 
we performed molecular dynamics (MD) simulations at room temperature using the DFTB method and calculated the effective conductance along the MD trajectories.
%
%
Strictly speaking, the MD trajectory should be calculated under non-equilibrium condition and the velocity for the MD should include the effect of  the heat produced tunneling current, however, we took the at equilibrium condition ignoring the heat produced by tunneling current for simplicity. At least from this calculation it is possible to examine the change of conductance due to the change of molecular structures as a crude approximation.

The motion of the atoms during the simulations has been integrated using  the standard Velocity Verlet algorithm with a time step of 1.0 fs and a total time duration of 10 ps. The influence of structural fluctuations on the conductance is investigated considering that only the atoms in the scattering region are allowed to move. 
After the MD simulation, 100 snapshots were extracted from the MD trajectories.
Then we calculated the effective conductance for the 100 stational geometries in order to see the change of conductance due to 
the change of the coordinate. The \textit{e-ph} coupling was set to the same as room temperature, 25.8 meV. \cite{Cattena}

\begin{figure}[ht!]
\begin{center}
\includegraphics[width=6cm,clip=true]{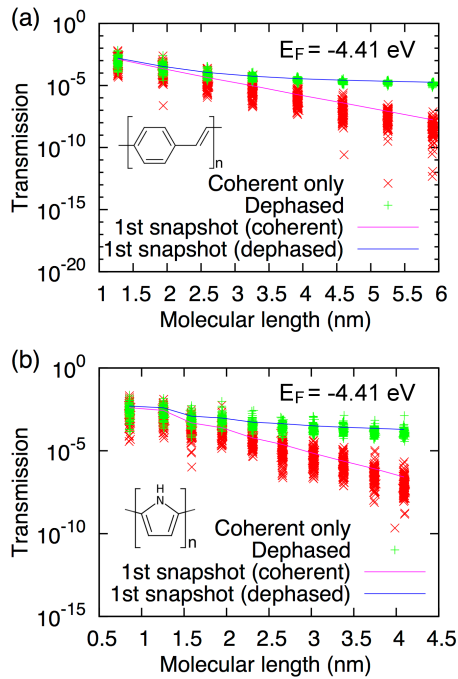}
\end{center}
\caption{\small{Electronic transmission at the Fermi energy within coherent and incoherent regimes as a function of length on the molecular dynamic flies for PPV-based (upper panel) and pyrrole-based molecular wires (lower panel). In the incoherent regime, dephasing strength was set to 25.8 meV.  }}
\label{fig5}
\end{figure}

\begin{figure}[ht!]
\begin{center}
\includegraphics[width=6.0cm,clip=true]{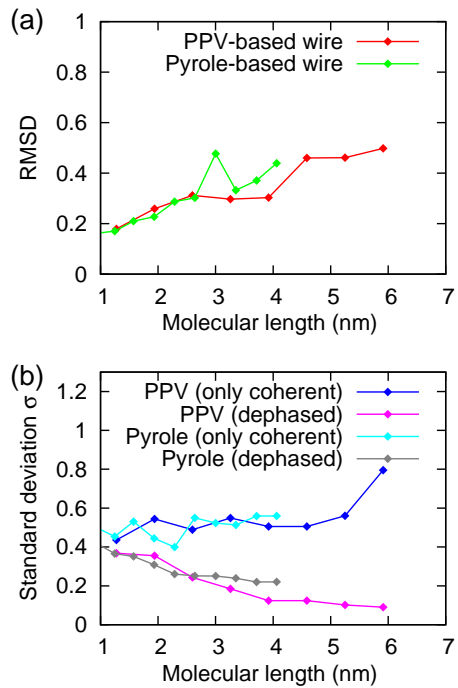}
\end{center}
\caption{\small{ (a) Fluctuation (RMSD)  of two types of molecular wires along MD pathways as a function of molecular lengths. (b) Standard deviation of $\log{T}$ for two types of molecular wires with/without dephasing effects. }}
\label{fig6}
\end{figure}

Figure \ref{fig5} shows how the coherent and effective conductances disperse along the MD trajectory with the molecular lengths. 
The dominance of hopping mechanisms for longer molecules is still preserved along the entire MD trajectory. The dephased transmission clearly deviates from the exponential trend of coherent tunneling. 

Another important feature between these two transport regimes is the degree of dispersion observed on the $\log{T}$ v.s.~length plots in Fig.~\ref{fig5}. 
Strong fluctuations are seen for the length dependence in the coherent tunneling regime whereas narrower distributions appear when  dephasing effects are considered. 
This issue can be clarified by analyzing the variation of the root-mean standard deviation (RMSD) of molecular coordinates with respect to a reference coordinate, 
in this case,  the initial one. 

Figure \ref{fig6} presents the molecular length dependence of RMSD (left panel) as well as the standard deviation (right panel) of the transmission logarithm ($\log{T}$) for the coherent and dephased regimes. RMSD results in Fig.~\ref{fig6}(a) point out that longer molecules potentially have more configurational degrees of freedom and, for this reason, RMSD increases with the size of the junction. Both types of molecules present approximately the same trends in the RMSD curves. 
The standard deviation of $\log{T}$ in Fig.~\ref{fig6}(b) highlights how   strongly the coherent/dephased transport fluctuates along MD trajectory. We can see that coherent tunneling is more sensitive to the molecular disorder induced by the thermal fluctuations than dephased transport. 
Thermal fluctuations can disrupt the $\pi$-orbital conjugation of the molecules, localizing 
the states along the molecular wires. Such breaking of the delocalized $\pi$-orbitals, which bridge the electron tunneling between two contacts, results in a large dispersion of coherent transmission in the $\log{T}$ v.s. length plots in Fig.~~\ref{fig5}. 

On the other hand, in the dephased regime,  unlike in the coherent regime, dephased transport does not
  fluctuate strongly with the breaking of $\pi$-orbital delocalization since there exists many individual jumps with probability $T_{m,n}$ for sequential hopping processes and there always will be pathways for hopping  that favor transport (note that dephased transport is determined by the sum of coherent tunneling and  sequential hoppings).  Therefore,   dephased transport is less sensitive to thermal fluctuations than coherent tunneling even if the $\pi$-orbital conjugation is thermally disturbed. Another important feature in the $\log{T}$ deviation within dephased transport is that it decreases for longer molecules. 
%
%
%
This is simply caused by the fact that the number of  paths through which the  charge  can be transmitted  increases as the molecule gets longer.  
The averaging over the increased number of  hopping pathways results in a smaller transmission dispersion of dephased transport. 

\section{\label{sec:5}Conclusion}
In summary, we have investigated electron transport in the presence of dephasing effects by analyzing both the coherent and incoherent process in molecular wires connected between two contacts. 
By employing the D'Amato-Pastawski model with first principle calculations, two transport mechanisms could be examined  under room temperature conditions where the structural conformations of the systems were explicitly treated.
We were able to observe the tunneling-to-hopping transition from exponential (coherent) to Ohmic (incoherent) regimes with molecular length using 
only one control parameter. 
We also investigated the influence of thermal fluctuations and found that the transmission distributions for tunneling transport are broader in comparison to those for the hopping mechanism. This highlights that the coherent component is rather more sensitive to structural disorder than the incoherent one. 
%
%
%
%
%
%
Our simple approach will provide a simulation platform for the interpretation of transport measurements in nanostructures.


%
%

\section*{Acknowledgments}
This work was funded by the Volkswagen Foundation and by the World  Class University program through the Korea Science and Engineering Foundation funded by the Ministry of Education, Science and Technology (Project No. R31-2008-000-10100-0), Erasmus Mundus, MPI-PKS, and Alexander von Humboldt Foundation.

\end{document}